\documentclass[12pt, amsfonts]{article}
\begin{document}
\def\p {{\partial}}
\def\n {{\nu}}
\def\m {{\mu}}
\def\a {{\alpha}}
\def\bt {{\beta}}
\def\f {{\phi}}
\def\th {{\theta}}
\def\g {{\gamma}}
\def\eps {{\epsilon}}
\def\e {{\psi}}
\def\k {{\chi}}
\def\la {{\lambda}}
\def\na {{\nabla}}
\def\bn {\begin{eqnarray}}
\def\en {\end{eqnarray}}
\title{The Hamilton-Jacobi treatment for an abelian Chern-Simons system\footnote{e-mail:
$sami_{-}muslih$@hotmail.com}} \maketitle
\begin{center}
\author{S. I. MUSLIH\\\it{Department of Physics, Al-Azhar
University,
Gaza, Palestine}}
\end{center}
\hskip 5 cm

$\mathbf{Abstract}$.- The abelian Chern-Simons system is treated
as a constrained system using the Hamilton-Jacobi approach. The
equations of motion are obtained as total differential equations
in many variables. It is shown that their simultaneous soultions
with the constraints lead to obtain canonical phase space
coordinates and the reduced phase space Hamiltonian with out
introducing Lagrange multipliers and with out any additional
gauge fixing condition.
\newpage

\section{Introduction}

Dirac method treatment [1, 2] for the abelian Chern-Simons field
interacting a scalar field in $2+1$ dimensions has been widely
discussed by many authors [3-10] in connection with particles
obeying fractional statistics or "anyons".

The presence of constraints in such theories require care when
applying Dirac's method, especially when first class constraints
arise, since the first class constraints are generators of gauge
transformations which lead to the gauge freedom and one should
impose external gauge fixing condition for each first class
constraint. Such gauge fixing is not always an easy task, for
example some authors [10] propose to use the coloumb gauge
$({\vec{\nabla}}{\cdot}{\vec{A}}=0)$ in order to obtain the
equations of motion and the correct reduced phase space
Hamiltonian for Chern-Simons system. Other authors [8,9] propose a
wrong gauge fixing $(A_{0}=0)$ to obtain the correct reduced
phase space coordinates.

In this paper we would like to use the canonical method [11-14] to
obtain the equations of motion and the correct reduced phase space
Hamiltonian for an abelian Chern-Simons field system interacting
with scalar field in $2+1$ dimensions. The equations of motion are
obtained as total differential equations in many variables. If
the system is integrable, then one can construct a valid and a
canonical phase space coordinates without using any gauge fixing
conditions.

\section{The Hamilton-Jacobi method of constrained systems}

In this section, we briefly review the Hamilton-Jacob method
[11-14] for studying the constrained systems.

This formulation leads us to obtain the set of Hamilton-Jacobi
partial differential equations [HJPDE] as follows:

\bn
&&H^{'}_{\a}(t_{\bt}, q_a, \frac{\p S}{\p q_a},\frac{\p S}{\p
t_a}) =0,\nonumber\\&&\a, \bt=0,n-r+1,...,n, a=1,...,n-r,\en where
\begin{equation}
H^{'}_{\a}=H_{\a}(t_{\bt}, q_a, p_a) + p_{\a},
\end{equation}
and $H_{0}$ is defined as
\bn
 &&H_{0}= p_{a}w_{a}+ p_{\m} \dot{q_{\m}}|_{p_{\n}=-H_{\n}}-
L(t, q_i, \dot{q_{\n}},
\dot{q_{a}}=w_a),\nonumber\\&&\m,~\n=n-r+1,...,n. \en

The equations of motion are obtained as total differential
equations in many variables as follows:

\bn
 &&dq_a=\frac{\p H^{'}_{\a}}{\p p_a}dt_{\a},\;
 dp_a= -\frac{\p H^{'}_{\a}}{\p q_a}dt_{\a},\;
dp_{\bt}= -\frac{\p H^{'}_{\a}}{\p t_{\bt}}dt_{\a}.\\
&& dz=(-H_{\a}+ p_a \frac{\p
H^{'}_{\a}}{\p p_a})dt_{\a};\\
&&\a, \bt=0,n-r+1,...,n, a=1,...,n-r\nonumber \en where
$z=S(t_{\a};q_a)$. The set of equations (4,5) is integrable
[11,12] if

\bn &&dH^{'}_{0}=0,\\
&&dH^{'}_{\m}=0,\;\; \m=n-p+1,...,n. \en If condition (6,7) are
not satisfied identically, one considers them as new constraints
and again testes the consistency conditions. Hence, the canonical
formulation leads to obtain the set of canonical phase space
coordinates $q_a$ and $p_a$ as functions of $t_{\a}$, besides the
canonical action integral is obtained in terms of the canonical
coordinates.The Hamiltonians $H^{'}_{\a}$ are considered as the
infinitesimal generators of canonical transformations given by
parameters $t_{\a}$ respectively.

\section{ An abelian Chern-Simons as a constrained system}

We consider an abelian Chern-Simons system as an example of the
applications of Hamilton-Jacobi procedure to treat constrained
systems. The Lagrangian density for Chern-Simons field
interacting a scalar field in $2+1$ dimensions is given by

\begin{equation}
{\cal {L}}= (D_{\m}\f)^{*}(D^{\m}\f) -{m}^{2}{\f}^{*}{\f}+
\frac{\kappa}{2}{\epsilon}_{\m\n\sigma}A^{\m}\p^{\n}A^{\sigma},
\end{equation}
where
\begin{equation}
D_{\m}= \p_{\m} -i eA_{\m}.
\end{equation}
The canonical momenta are defined as \bn&& \pi_{\m}= \frac{\cal
L}{\p \dot{A^{\m}}}= \frac{1}{2}\kappa {\epsilon}_{0\m\rho}A^{\rho},\\
&&p_{\f}= \frac{\cal L}{\p \dot{\f}}= (D_{0}\f)^{*} =\dot{\f^{*}}
+i e A_{0}{\f}^{*},\\
&&p_{{\f}^{*}}= \frac{\cal L}{\p \dot{\f^{*}}}= (D_{0}\f)
=\dot{\f} - i e A_{0}{\f}. \en The non vanishing Poisson brackets
are \bn&&\{{\f}(\vec{x}, t), p_{\f}(\vec{y}, t)\}=\delta(\vec{x}-
\vec{y}),\\
&&\{{\f}^{*}(\vec{x}, t), p_{{\f}^{*}}(\vec{y},
t)\}=\delta(\vec{x}-
\vec{y}),\\
&&\{{A^{\m}}(\vec{x}, t), \pi_{\n}(\vec{y},
t)\}={\delta}_{\n}^{\m}\delta(\vec{x}- \vec{y}). \en Upon
quantization, these brackets have to be converted into proper
commutators.

 The canonical momenta $\pi_{\m}$ conjugated to $A_{\m}$ lead
us
to the primary constraints \bn&&{H'}_{1}= \pi_{0}=0,\\
&&{H'}_{2}= \pi_{1} - \frac{1}{2} \kappa A^{2}=0,\\
&&{H'}_{3}= \pi_{2} + \frac{1}{2} \kappa A^{1}=0. \en The
canonical Hamiltonian $H_{0}$ can be written as \bn H_{0}= &&\int
d^{2}x[p_{\f^{*}}p_{\f} + A_{0}(\kappa
{\vec{\nabla}}{\times}{\vec {A}} - j_{0}) +
{\vec{\nabla}}{\f}^{*}{\cdot}{\vec{\nabla}}{\f} \\ &&- ie
{\vec{A}}{\cdot}({\f}^{*}{\nabla^{\leftrightarrow}} {\f}) +e^{2}
|{\vec{A}}|^{2}|{\f}|^{2} + m^{2}|{\f}|^{2}], \en where
\begin{equation}
{\vec{A}}=(A^{1}, A^{2}),\;\;\;{\vec{\nabla}}=(\p_{1},
\p_{2}),\;\;\;\; {\vec{\nabla}}{\times}{\vec{A}}=(\p_{1}A^{2} -
\p_{2}A^{1}),
\end{equation}
and
\begin{equation}
j^{0}= ie(p_{\f^{*}}{\f}^{*}- p_{\f}{\f} )= ie({\f}^{*}
D_{0}{\f}- (D_{0}\f)^{*}{\f}).
\end{equation}

Now the canonical method leads us to obtain the set of
Hamilton-Jacobi partial differential equations as: \bn&&{H'}_{0}=
\pi_{4} + H_{0}=0;\;\;\;\pi_{4}= \frac{\p S}{\p x_{0}},\\
&&{H'}_{1}= \pi_{0}=0,\;\;\;\;\;\;\;\;\;\;\;\;\;\;\;\;\;\;\;\pi_{0}= \frac{\p S}{\p A^{0}},\\
&&{H'}_{2}= \pi_{1} - \frac{1}{2} \kappa A^{2}=0,\;\;\;\;\pi_{1}= \frac{\p S}{\p A^{1}},\\
&&{H'}_{3}= \pi_{2} + \frac{1}{2} \kappa
A^{1}=0,\;\;\;\;\;\pi_{2}= \frac{\p S}{\p A^{2 }}. \en

The equations of motion are obtained as total differential
equations in many variables as follows: \bn d{\f}&&= \frac{\p
{H'}_{0}}{\p p_{\f}}dx^{0} + \frac{\p {H'}_{1}}{\p p_{\f}}dA^{0} +
\frac{\p {H'}_{2}}{\p p_{\f}}dA^{1} + \frac{\p {H'}_{3}}{\p
p_{\f}}dA^{2},\nonumber\\
&&= (p_{{\f}^{*}}+ i eA_{0}{\f})dx^{0},\\
d{\f}^{*}&&= \frac{\p {H'}_{0}}{\p p_{{\f}^{*}}}dx^{0} + \frac{\p
{H'}_{1}}{\p p_{{\f}^{*}}}dA^{0} + \frac{\p {H'}_{2}}{\p
p_{{\f}^{*}}}dA^{1} + \frac{\p {H'}_{3}}{\p
p_{{\f}^{*}}}dA^{2},\nonumber\\
&&= (p_{{\f}}- i eA_{0}{\f}^{*})dx^{0},\\
dp_{\f}&&= -\frac{\p {H'}_{0}}{\p {\f}}dx^{0} - \frac{\p
{H'}_{1}}{\p {\f}}dA^{0} - \frac{\p {H'}_{2}}{\p {\f}}dA^{1} -
\frac{\p {H'}_{3}}{\p
{\f}}dA^{2},\nonumber\\
&&= (({\vec{D}}{\cdot}{\vec{D}}\f)^{*}- m^{2}{\f}^{*} -ieA_{0} p_{\f})dx^{0},\\
dp_{{\f}^{*}}&&= -\frac{\p {H'}_{0}}{\p {\f}^{*}}dx^{0} - \frac{\p
{H'}_{1}}{\p {\f}^{*}}dA^{0} - \frac{\p {H'}_{2}}{\p
{\f}^{*}}dA^{1} - \frac{\p {H'}_{3}}{\p
{\f}^{*}}dA^{2},\nonumber\\
&&= (({\vec{D}}{\cdot}{\vec{D}}\f)- m^{2}\f + ieA_{0} p_{{\f}^{*}})dx^{0},\\
d\pi_{0}&&= -\frac{\p {H'}_{0}}{\p {A}^{0}}dx^{0} - \frac{\p
{H'}_{1}}{\p {A}^{0}}dA^{0} - \frac{\p {H'}_{2}}{\p
{A}^{0}}dA^{1} - \frac{\p {H'}_{3}}{\p
{A}^{0}}dA^{2},\nonumber\\
&&= -(\kappa {\vec{\nabla}}{\times}{\vec {A}} - j_{0})dx^{0},\\
d\pi_{1}&&= -\frac{\p {H'}_{0}}{\p {A}^{1}}dx^{0} - \frac{\p
{H'}_{1}}{\p {A}^{1}}dA^{0} - \frac{\p {H'}_{2}}{\p {A}^{1}}dA^{1}
- \frac{\p {H'}_{3}}{\p
{A}^{1}}dA^{2},\nonumber\\
&&= (-\kappa \p_{2}A^{0} + j_{1})dx^{0}- \frac{1}{2} \kappa dA^{2},\\
d\pi_{2}&&= -\frac{\p {H'}_{0}}{\p {A}^{2}}dx^{0} - \frac{\p
{H'}_{1}}{\p {A}^{2}}dA^{0} - \frac{\p {H'}_{2}}{\p
{A}^{2}}dA^{1} - \frac{\p {H'}_{3}}{\p
{A}^{2}}dA^{2},\nonumber\\
&&= (\kappa \p_{1}A^{0} + j_{2})dx^{0} + \frac{1}{2} \kappa
dA^{1}, \en where
\begin{equation}
j_{i}= ie({\f}^{*}{{\p_{i}}^{\leftrightarrow}}{\f} -2e^{2}
A^{i}|\f|^{2}=ie({\f}^{*}D_{i}{\f} -(D_{i}{\f})^{*}{\f}),
\end{equation}
are the space components of the current $j_{\m}=
ie({\f}^{*}{D_{\m}}^{\leftrightarrow}{\f}$ and we use the notation
${\f}^{*}D_{\m}= (D_{\m}{\f})^{*}$.

To check whether the set of equations (27-33) is integrable or
not, we have to consider the total variations of the constraints.
In fact
\bn&&d{H'}_{2}=(-\kappa \p_{2}A^{0} + j_{1})dx^{0}- \kappa dA^{2}=0,\\
&&d{H'}_{3}=(\kappa \p_{1}A^{0} + j_{2})dx^{0} + \kappa dA^{1}=0.
\en

The constraints (25,26), lead us to obtain $dA^{1}$ and $dA^{2}$
in terms of $dt$ \bn&&dA^{1}=(- \p_{1}A^{0} -\frac
{j_{2}}{\kappa})dx^{0},\\ &&dA^{2} =(- \p_{2}A^{0} +
\frac{j_{1}}{\kappa})dx^{0}. \en

The vanishing of the total differential of ${H'}_{1}$ leads to a
new constraint
\begin{equation}
{H'}_{4}= (\kappa {\vec{\nabla}}{\times}{\vec {A}} - j_{0}).
\end{equation}
Taking the total differential of ${H'}_{4}$ leads to the
conservation of current
\begin{equation}
\p_{\m}j^{\m}=0.
\end{equation}

The set of equations (27-33) is integrable. Hence, the canonical
phase space coordinates $(\f, p_{\f})$ and $({\f}^{*},
p_{{\f}^{*}})$ are obtained in terms of parameters $(t, A^{0})$.
Besides, the reduced phase-space Hamiltonian can be calculated as
\begin{equation}
H_{0}=\int d^{2}x[p_{\f^{*}}p_{\f}
+{{\vec{D}}{\f}}^{*}{\cdot}{\vec{D}}{\f}
 + m^{2}|{\f}|^{2}],
 \end{equation}
 where
 \begin{equation}
 {\vec{D}}={\vec{\nabla}} +ie {\vec{A}},
 \end{equation}
\section{ Conclusion}

The basic idea of Dirac's method to investigate an abelian
Chern-Simons system is to consider the total Hamiltonian composed
by adding the constraints multiplied by Lagrange multipliers to
the canonical Hamiltonian. In order to derive the equations of
motion, one needs to redefine these unknown multipliers in an
arbitrary way [8-10]. However, in the Hamilton-Jacobi method
[11-14], there is no need to introduce Lagrange multipliers to the
canonical Hamiltoian.

In this paper we have obtained the canonical phase space
coordinates and the reduced canonical Hamiltonian for an abelian
Chern-Simons system with out using any gauge fixing condition.


\begin{thebibliography}{widest-label}
\bibitem{1}DIRAC P. A. M., {\it Lectures on Quantum Mechanics}, Belfer
 Graduate School of Science, Yehiva University (A cademic Press,
 New York)1964.
\bibitem{2}DIRAC P. A. M., Can. J. Math., $\mathbf{2}$ (1950)129.
\bibitem{3}PANERJEE R., Phys. Rev. D $\mathbf{48}$ (1993) 2905.
\bibitem{4}HAGEN C., Ann. Phys. (N. Y.) $\mathbf{157}$ (1984) 342.
\bibitem{5} FORTE S., Rev. Mod. Phys. $\mathbf{64}$ (1993) 193.
\bibitem{6}FERRARI F and LAZZIZZERA I. HEP-TH/9611211.
\bibitem{7}FORESTER A. and GIROTTI H. O., Phys. Lett. $\mathbf{B
 203}$(1989) 83.
\bibitem{8}SEMENOFF G., Phys. Rev. Lett. $\mathbf{61}$ (1984) 342.
\bibitem{9}SEMENOFF G. and SODANO P., Nucl. Phys. $\mathbf{B  328}
$(1989) 753.
\bibitem{10}SHARAN P., MEHRA A. and DASGUPTA K., HEP-TH/9903098.
\bibitem{11}MUSLIH S. I. and GULER Y., Nuovo Cimento B, $\mathbf{110}$
(1995) 307.
\bibitem{12}MUSLIH S. I. and GULER Y., Nuovo Cimento B, $\mathbf{113}$
(1998)277
\bibitem{13}GULER Y., Nuovo Cimento B, $\mathbf{107 }$(1992)1389.
\bibitem{14}GULER Y., Nuovo Cimento B, $\mathbf{107}$ (1992)1143.

\end{thebibliography}
\end{document}